\PassOptionsToPackage{numbers,sort&compress,square}{natbib}

\documentclass[floatfix,%
 aip,
 amsmath,amssymb,
 reprint,%
]{revtex4-1}

\usepackage{graphicx}
\usepackage{dcolumn}
\usepackage{bm}

\usepackage[utf8]{inputenc}
\usepackage[T1]{fontenc}
\usepackage{mathptmx}
\usepackage{etoolbox}
\usepackage{siunitx}
\usepackage[numbers,sort&compress,square]{natbib}
\setcitestyle{super,sort&compress}

\usepackage{booktabs} 

\usepackage[none]{hyphenat}
\makeatletter
\def\@email#1#2{%
 \endgroup
 \patchcmd{\titleblock@produce}
  {\frontmatter@RRAPformat}
  {\frontmatter@RRAPformat{\produce@RRAP{*#1\href{mailto:#2}{#2}}}\frontmatter@RRAPformat}
  {}{}
}%
\makeatother
\begin{document}

\preprint{AIP/123-QED}

\title[]{A cryogenic apparatus with a rotatable substrate for matrix isolation spectroscopy of BaF in neon crystals down to 3.2 K}

\author{Alessandro Lippi *}
\affiliation{Istituto Nazionale di Fisica Nucleare, Sezione di Ferrara, 44122 Ferrara, Italy}
\affiliation{Department of Physics and Earth Science, University of Ferrara, 44122 Ferrara, Italy}
\email{lpplsn@unife.it}

\author{Diego Bisero}
\affiliation{Istituto Nazionale di Fisica Nucleare, Sezione di Ferrara, 44122 Ferrara, Italy}
\affiliation{Department of Physics and Earth Science, University of Ferrara, 44122 Ferrara, Italy}

\author{Caterina Braggio}
\affiliation{Istituto Nazionale di Fisica Nucleare, Sezione di Padova, 35131 Padua, Italy}
\affiliation{Department of Physics and Astronomy, University of Padua, 35131 Padua, Italy}

\author{Giovanni Carugno}
\affiliation{Istituto Nazionale di Fisica Nucleare, Sezione di Padova, 35131 Padua, Italy}
\affiliation{National Quantum Science and Technology Institute, 00185 Rome, Italy}

\author{Roberto Calabrese}
\affiliation{Istituto Nazionale di Fisica Nucleare, Sezione di Ferrara, 44122 Ferrara, Italy}
\affiliation{Department of Physics and Earth Science, University of Ferrara, 44122 Ferrara, Italy}

\author{Federico Chiossi}
\affiliation{Istituto Nazionale di Fisica Nucleare, Sezione di Padova, 35131 Padua, Italy}
\affiliation{Department of Physics and Astronomy, University of Padua, 35131 Padua, Italy}

\author{Marco Guarise}
\affiliation{Istituto Nazionale di Fisica Nucleare, Sezione di Ferrara, 44122 Ferrara, Italy}
\affiliation{Department of Physics and Earth Science, University of Ferrara, 44122 Ferrara, Italy}

\author{Madiha M. Makhdoom}
\affiliation{Istituto Nazionale di Fisica Nucleare, Sezione di Padova, 35131 Padua, Italy}
\affiliation{Department of Physics and Astronomy, University of Padua, 35131 Padua, Italy}
\affiliation{Interdisciplinary Research Center for Advanced Materials, King Fahd University of Petroleum and Minerals, Dhahran, 31261, Saudi Arabia}

\author{Giuseppe Messineo}
\affiliation{Istituto Nazionale di Fisica Nucleare, Sezione di Padova, 35131 Padua, Italy}
\affiliation{National Quantum Science and Technology Institute, 00185 Rome, Italy}

\author{Jacopo Pazzini}
\affiliation{Istituto Nazionale di Fisica Nucleare, Sezione di Padova, 35131 Padua, Italy}
\affiliation{Department of Physics and Astronomy, University of Padua, 35131 Padua, Italy}

\collaboration{DOCET}

\date{\today}

\begin{abstract}

We present a new experimental apparatus for matrix isolation spectroscopy capable of growing cryogenic crystal matrices doped with barium monofluoride (BaF), a candidate for measurements of the electron electric dipole moment (eEDM). 
\\
In this work, BaF molecules are produced in a cryogenic buffer-gas source and used to dope a neon matrix grown on a substrate, cooled by a cryocooler, reaching temperatures as low as 3.2 K. A rotatable flange allows for multiple substrate orientations, optimizing spectroscopic investigations and matrix crystal growth.
\\
Fluorescence emission analysis of BaF in neon crystal show emission when exciting the $A\,^2\Pi_{1/2}$, $A\,^2\Pi_{3/2}$ and  $B\,^2\Sigma^+$ states, along with the presence of fluorescence from barium atoms in the crystal. A first estimation of the column density of BaF in the crystal is given on the order of $5 \times 10^{-13} \, \text{cm}^{-2}$.
\\
We also demonstrated the production of BaF using parahydrogen as the buffer gas of the cryogenic source, a first step toward BaF molecules embedded in a para-hydrogen matrix.  
\\
Finally, we also observed fast fluorescence decay, attributed to optical bleaching effects, whose amplitude arise at lower temperatures or higher laser intensities, introducing a systematic effect to be studied in future experiments.

\end{abstract}

\maketitle

\section{\label{sec:level1} Introduction}

Matrix Isolation Spectroscopy (MIS) is a well established technique for studying reactive atoms and molecules by inhibiting their motion within a minimally interacting environment\,\cite{barnes2012matrix,fajardo2011physics}. By embedding guest species, such as atoms, ions, or molecules within a rigid, chemically inert lattice (typically a solid rare gas matrix) at cryogenic temperatures, MIS effectively freezes their movements. This allows for an in-depth investigation of the guest species. Crucially, by minimizing intermolecular interactions between the host and guest, this method provides gas-like spectrum while the sample is held in the solid phase.
Nowadays MIS is a widely used experimental technique in many fields in chemistry\,\cite{momose1998matrix,moore2021hydrocarboxyl,bahou2014infrared}, astrochemistry\,\cite{feldman2021matrix,weber2024unique} and fundamental physics\,\cite{rollings2026long,messineo24, lahs2026atomic,guarise2022particle,guarise2020particle}.

The most common preparation method for doped matrices, involves the co-deposition of a highly diluted gaseous mixture, consisting of the target guest and an inert host, onto a cryogenically cooled substrate\,\cite{guarise2017experimental}. A primary challenge of MIS is maintaining a sufficiently low temperature to prevent the host matrix from sublimating or allowing the trapped molecules to diffuse and recombine. Therefore, highly controlled systems are essential to this field. These systems must provide reliable access to cryogenic temperatures, stable operation across various materials, and multiple optical access points to facilitate comprehensive spectroscopic study. Moreover, MIS experiment require repeated spectroscopic characterization with multiple optical geometries, magnetic field access and the lowest temperatures achievable, all difficult requirements in conventional fixed-geometry cryogenic setups.
\\
In this paper we address these limitations, as we developed an apparatus for the growth of cryogenic matrices such as neon or para-hydrogen (p-H$_2$) for MIS, combining independent cryogenic cooling and a rotatable flange allowing for multiple spectroscopic analysis. 
\\
Cryogenic matrices will be doped with diatomic polar molecules such as monofluoride (BaF) for fundamental physics researches of the electron electric dipole moment (eEDM). Such cryogenic systems will in fact permit the investigation of a large number of molecules in a confined environment allowing to perform precise study of the behavior of electrons in high electric fields such as the ones provided by BaF molecules.

\section{Experimental setup}

The experimental setup designed to produce the doped cryogenic matrix is based on a six-way cross stainless-steel ultra-high-vacuum (UHV) chamber. The four horizontal ports are equipped with optical viewports for laser spectroscopy and visual inspection whereas on the top and bottom port two cryocoolers are installed. The first, a Sumitomo RP62B, is used to cool down to 3.2 K a sapphire substrate mounted on a rotating flange, while the second, a Leybold-Heraeus RG210, serves as cooling system for the ablation cell mounted on a movable vertical stage. 
\\
The UHV is maintained through an additional four-way cross mounted between the main chamber and the Sumitomo cryocooler on top. The Sumitomo is mounted on a rotatable flange, a key feature of this apparatus as it allows for multiple geometries for optical investigations. The four-way cross hosts both a turbo-molecular pump TV 551, with a backing pump, that reaches a vacuum of $1\times 10 ^{-7}$ mBar, and vacuum and thermal diagnostic. A Pfeiffer cold-cathode ionization gauge PKR 251 is used to measure the vacuum pressure. Cernox sensors and resistive heaters, connected through electrical feedthroughs, are also installed in the same flange and positioned at the top of the two cold fingers, allowing continuous monitoring of the temperatures and independent heating of the two cryo-stages. 

\subsection*{Molecule production}
\begin{figure}[]
\centering
\includegraphics[width=1\linewidth]{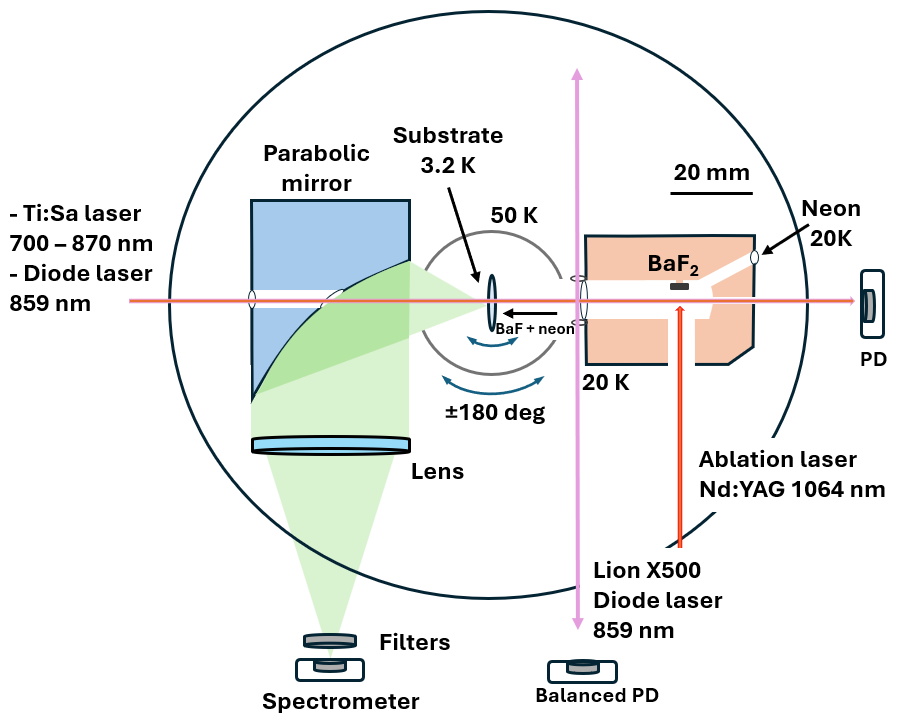}
\caption{Schematic of the experimental setup for the production and detection of BaF molecules by laser ablation in a cryogenic buffer-gas cell. A pulsed 1064\,nm laser ablates a BaF$_2$ target inside the copper cell. The produced molecules are cooled by the buffer gas and extracted through a 3\,mm aperture toward the cryogenic substrate, where the matrix is grown. A diode laser tuned to the $A\,^2\Pi_{1/2}$ transition at 859\,nm monitors the production of BaF molecules both inside the ablation cell and after extraction. The substrate is mounted on a rotatable cold finger, allowing it to face either the ablation, during matrix deposition, or the fluorescence collection optics, during spectroscopic measurements. In the latter configuration, the excitation laser directly addresses the matrix while the emitted fluorescence is collected by a parabolic mirror and detected with either a spectrometer or a photodiode.}
\label{fig:setup}
\end{figure}

Molecules are produced by ablation of a solid target inside the ablation cell with the buffer gas cooling technique \cite{Hutzler_buffergas}. The cell consists of a machined copper block with outside dimensions of ($40 \times 30 \times 30)~\mathrm{mm}^3$. It incorporates two optical access ports with diameters 10\,mm and 3\,mm, respectively, sealed with CaF$_2$ windows, providing optical access for ablation and absorption lasers. A third aperture, 3\,mm in diameter, serves as the molecular beam extraction hole towards the cryogenic substrate and simultaneously provides optical access to the interior of the ablation cell. At the center of the copper block a cylindrical hole, with 15\,mm diameter and 30\,mm depth, shown in Figure \ref{fig:setup}, houses the solid precursor target of BaF$_2$, which is mounted directly opposite to the 10\,mm optical port used for the ablation laser. Molecules are produced by pulsed laser ablation of the target through a pulsed 1064\,nm EKSPLA NL300 Nd:YAG laser, with a repetition rate of 10\,Hz and energies ranging between 10\,mJ and 30\,mJ. The laser light is focused on the BaF$_2$ target in a spot with a diameter of approximately 0.5\,mm, producing a plasma plume that evaporates the desired atoms or molecules, producing Ba, BaF$_2$ and BaF. \\
To probe and enhance the production of BaF molecules in the gas phase, two absorption measurements inside and outside of the cell are performed simultaneously at 859\,nm on the $X^2\,\Sigma^+\,(v\!=\!0) $ to $ A^2\,\Pi_{1/2}\,(v\!=\!0)$ transition, providing a parameter to control molecule production. For these measurements, an external-cavity diode laser (ECDL) tunable near 859\,nm (Lion X500, Sacher GmbH) is used. A similar absorption scheme is described in our previous work \cite{lippi2025neutralbariumsolidneon}.
\\
The ablation spot is periodically shifted to maintain a stable production, as heating up the target reduces the ablation efficiency. Moreover, leaving the ablation spot on the same position produces holes in the target.
\\
A buffer gas, generally Ne or p-H$_2$ in our experiments, is used to cool down and carry out the products from the cell extraction hole with an expected velocity of 210\,m/s \cite{maarten2025a}. 
The buffer gas is inserted into the ablation cell through a cylindrical inlet oriented at an angle with respect to the molecular beam exit aperture, as shown in Figure \ref{fig:setup}, directing the buffer gas towards the plasma plume and the extraction hole to optimize its exit from the cell. The ablation cell is surrounded by a 50\,K shield and reaches a base temperature of 16\,K. If neon is used to cool down the products, a flow is maintained between 10 to 20\,SCCM while the cell is heated with a resistor at approximately 20\,K, to suppress unwanted condensation of the gas, while still allowing efficient cooling of the molecules.
\\
In our experiment, the BaF$_2$ target is prepared under vacuum conditions and baked at high temperature to minimize contaminants that would otherwise reduce molecular yield. 
\\
Production of BaF in the gas phase using barium as the ablation target and implementing a second gas in the ablation cell, such as CF$_4$ or SF$_6$, is known to have a higher yield of BaF molecules in the production process\cite{van20262d}. However, extra gases in the system will result in additional dopants onto the final cryogenic matrix. 
\\
BaF molecules formed by laser ablation are entrained in a continuous neon flow and delivered to the substrate for matrix growth. A neon crystal is usually produced for 5 to 10 minutes before the ablation process. This initial layer prevents the ablation products with high sublimation temperatures from being deposited directly onto the sapphire, allowing efficient sublimation of all compounds when the crystal is heated up to 40\,K.
\\
The ablation cell extraction hole is located at 30\,mm from the substrate during the production of molecules and crystal growth. Once the crystal is produced, the ablation cell can be lowered by an edge welded bellow of about 50\,mm, clearing the optical axis for fluorescence measurements.  
\\
Molecule production was further enhanced by orienting the surface of the  BaF$_2$ pellet at an angle, directing the ablation plume toward the extraction hole of the cell.


\subsection{Preliminary BaF production using para-hydrogen buffer gas }
\begin{figure}[]
\centering
\includegraphics[width=1\linewidth]{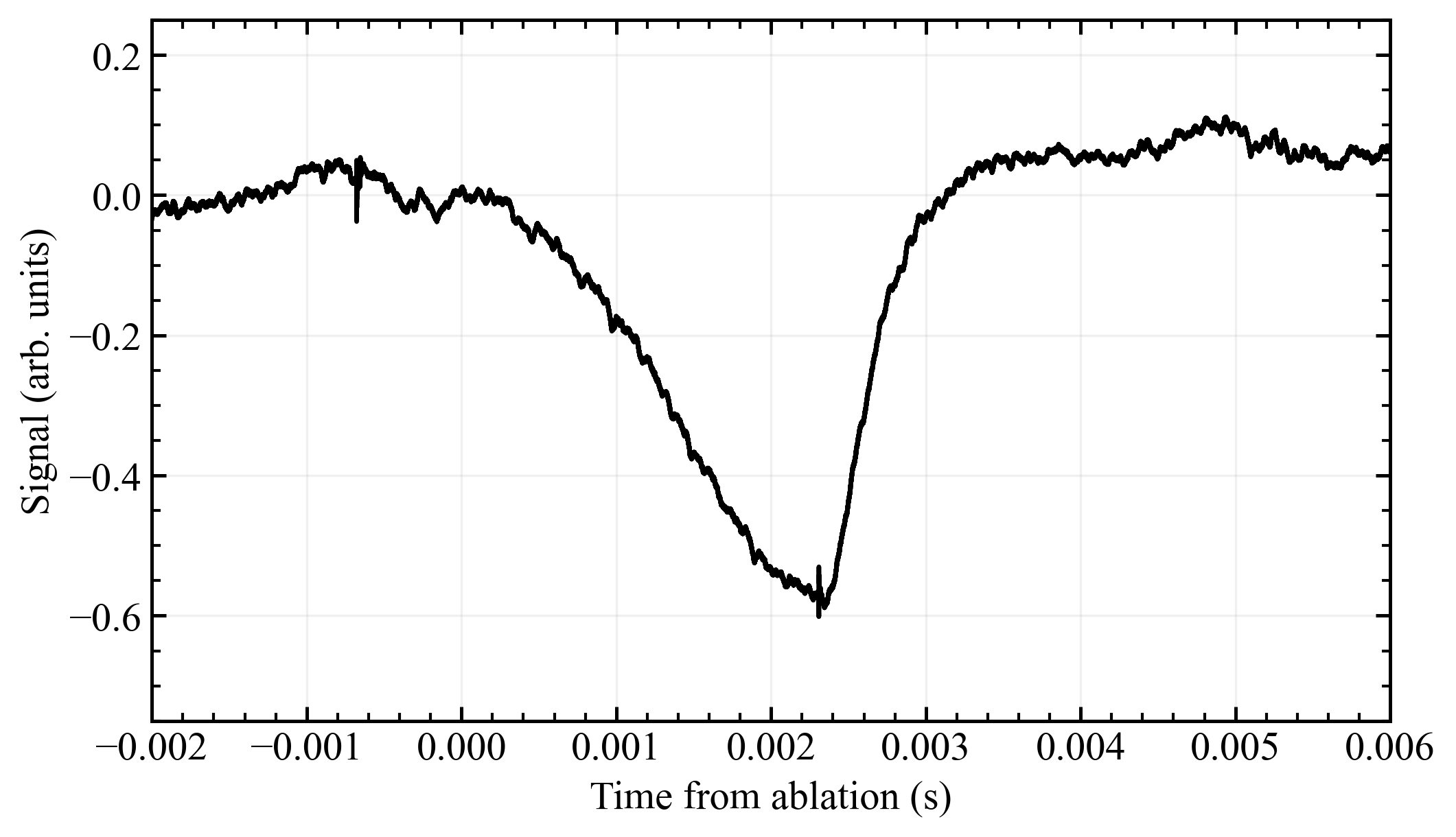}
\caption{Absorption signal of BaF molecules produced inside the cryogenic buffer-gas cell using p-H$_2$ as the buffer gas. The trace corresponds to laser absorption on the $X^2\,\Sigma^+\,(v=0)$ to $ A^2\,\Pi_{1/2}\,(v'=0)$ transition at 11630.50 cm$^{-1}$ (859\,nm). The signal was achieved with a p-H$_2$ flux of approximately 40\,SCCM and an ablation energy of 25\,mJ. }
\label{fig:ass_pH}
\end{figure}

The apparatus has been designed to operate and condense p-H$_2$, a promising host matrix for future precision measurements on polar molecules \cite{rollings2026long, weinstein16, weinstein20}.
\\
In our previous work we have proven to be able to produce p-H$_2$ with 2\% of ortho-hydrogen \cite{parahydrogen_us}. Since then, a purer p-H$_2$ production has been achieved thanks to a lower temperature in the ortho to para conversion chamber, going from 25\,K down to 15\,K\,\cite{tom2009producing}.
\\
As a first, to produce a doped p-H$_2$ matrix, we present a preliminary result of production of BaF molecules using p-H$_2$ as a buffer gas. Figure \ref{fig:ass_pH} shows a representative absorption signal measured inside the source, corresponding to the transition from the ground state $X^2\,\Sigma^+(v=0)$ to $ A^2\,\Pi_{1/2}(v'=0)$.
\\
Compared with noble gases such as neon or argon, molecular hydrogen thermalizes heavy species less efficiently, as its smaller mass requires a larger number of collisions to reach thermal equilibrium. The lower mass also increases the diffusion of BaF, enhancing the losses to the cell walls. Consequently, larger gas flows were required to obtain a detectable molecular signal. To achieve our best result of absorption inside the ablation cell of 10\% (0.6 V over 6 V on a photodiode) we reached the maximum regime of our fluxmeter, approximately 40 SCCM, more than two times the flux used with neon as a buffer gas.
\\
Although these measurements are preliminary, they demonstrate that the present source is compatible with a production of BaF molecules using p-H$_2$, the first step for our future works on BaF embedded in p-H$_2$ crystals.

\subsection*{Cryogenic crystal growing setup}
The cryogenic matrix is grown on a 1/2" (12\,mm) substrate, either a sapphire or a CaF$_2$ window, depending on the needs of the experiment. A sapphire substrate is known to have better thermal conductivity, however, CaF$_2$ has better optical properties and does not emit fluorescence light in the spectral range of this work. \\
The substrate is clamped to a copper holder anchored to the cold finger of the Sumitomo two-stage cryocooler. A Cernox sensor, a thin-film resistance temperature detector, is placed at 5\,cm from the substrate to measure its temperature. The lowest temperature recorded is 3.2 K, whereas under standard operating conditions the substrate remains at an average temperature of 3.5 K and reaches a temperature of 4 K when the buffer gas, at a standard temperature of 20 K, condenses on it. In order to reach such low temperatures, a 50 K radiative shield surrounds the cold finger, with the exception of two 15\,mm holes that allow for gas deposition and an optical path over the crystal. Close to the Cernox sensor, a heater is positioned to heat the sample, used for annealing or sublimation processes.
\\ 
The substrate holder is mounted on a rotatable flange featuring a dual viton seal with an insulation vacuum. The flange allows for 180$^\circ$ rotation around its vertical axis, a key feature of the apparatus as it enables the orientation of the substrate to be optimized depending on the stage of the experiment, allowing multiple matrix diagnostic during the same crystal growth, minimizing systematic effects during crystal characterization as it eliminates the need for repeated crystal growth. 
\\
During laser ablation and crystal growth, the substrate faces the ablation cell. Rotating the substrate then gives access to two complementary geometries on the very same crystal, without the need for repeated growths.
\\
For fluorescence measurements, the substrate is rotated by 180$^\circ$, as shown in Figure \ref{fig:setup}, facing a parabolic collection mirror with an effective focal length of $f_{\mathrm{e}} = 46$\,mm. In this configuration the excitation laser addresses directly the BaF molecules in the crystal without passing through either the ablation cell or the substrate, avoiding possible scattering, birefringence, and background fluorescence effects.
\\
An intermediate 90$^\circ$ orientation instead provides the geometry required for transmission and absorption spectroscopy, enabling characterization of the cryogenic matrix, particularly p-H$_2$ crystals, following the methodology of previous works with FT-IR interferometry \cite{fajardo2011physics, parahydrogen_us, weinstein21}.
\\
Because all three configurations, ablation, fluorescence, and transmission, are accessible on a single crystal growth, the same sample can be diagnosed with multiple complementary techniques while eliminating systematic effects arising from the growth of a new crystal for each measurement.
\\
A set of external coils is also implemented in the setup to generate a magnetic field of approximately 30 G perpendicular to the substrate, available in the fluorescence configuration.

\subsection*{Fluorescence collection setup} 
Excitation was performed using two laser systems, depending on the measurement requirements: an external-cavity diode laser with output power up to 30 mW between 830--870\,nm and an Avesta laser with a linewidth of approximately 2\,GHz, tunable over the 700--900\,nm range and capable of output powers ranging from 10 to 300\,mW. \\
The excitation beam passes through the parabola aperture and is focused on a spot diameter of 0.7\,mm on the crystal, reaching a maximum power density onto the crystal of about 100\,W/cm$^2$ when pulsed through an acousto-optic modulator. 
\\
As mentioned above, the fluorescence light collection happens through a light collecting parabola focusing on the crystal. One lens is mounted in the vacuum chamber on the same support of the parabolic mirror and it is used to focus the light emitted from the crystal outside of the chamber.
\\
The fluorescence detection apparatus is composed of interchangeable detection modules. For high range spectroscopy we use the Spectral Products SM642 spectrometer for the visible range and Ocean Optics NIR Quest InGaAs for near-infrared light. An InGaAs detector is otherwise used along side a Thorlabs PDA. The modular detection scheme allows for rapid switching between the two types of measurements without modification of optical alignment onto the crystal.

\section{barium monofluoride detection in a neon matrix at 3.2\,K}
\begin{figure}[]
\centering
\includegraphics[width=1\linewidth]{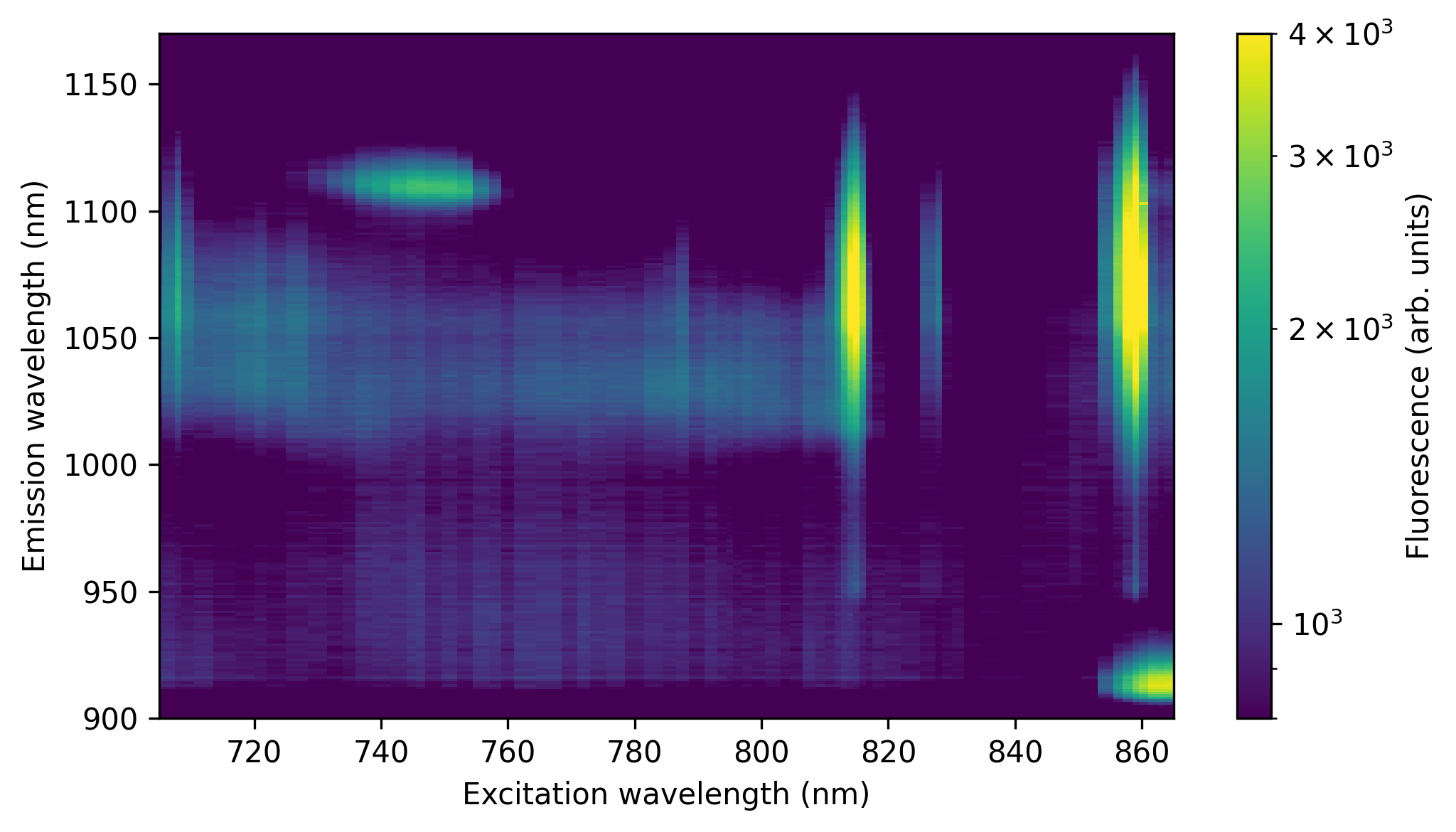}
\caption{2D excitation–emission spectrum of BaF embedded in a neon matrix at 3.2\,K, recorded using a single excitation laser with a power of 45\,mW focused onto a spot diameter of approximately 0.7\,mm. The main excitation bands associated with the $A\,^2\Pi_{1/2}$ (859\,nm) and $A\,^2\Pi_{3/2}$ (815\,nm) electronic states, together with their first vibrational transitions, are observed, as well as fluorescence from the $B\,^2\Sigma^+$ state (708\,nm). Additional emission features are also detected. A broad band excited around 750\,nm and emitting at 1107\,nm is attributed to neutral barium atoms trapped in the neon matrix. A second broad emission band, centered around 1045\,nm and excited over the wavelength range from approximately 815 to 700\,nm, is observed but remains unassigned.}
\label{fig:scan2D}
\end{figure}

With our setup, we performed an analysis similar to that of Li et al. \cite{Vutha2023}, but in this setup we are able to investigate in a different range of excitation and at a temperature of 3.2\,K demonstrating the spectroscopic capabilities of the apparatus. Results are shown in Figure \ref{fig:scan2D}, with a maximum excitation wavelength of 870\,nm and minima of 700\,nm. 
In alignment with previous works, the two main excitation bands from the ground state are detected with their respective first vibrational levels, mainly: 815\,nm and 787\,nm ($v=1$) for $ A\, ^2\Pi_{3/2}$; 859\,nm and 828\,nm ($v=1$) for $ A\,^2\Pi_{1/2}$.
\\
A third excitation band, previously reported in the work of Corriveau et al. \cite{Hessels2024}, to the $B\,^2\Sigma^+$ state (712\,nm), is shifted with respect to the gas phase by approximately 4\,nm (708\,nm) on the blue side.
\\
Excitation bands are, as in the work of Li et al. \cite{Vutha2023}, obtained from the integration of the fluorescence light between (950 - 970)\,nm or (1000 - 1100)\,nm, corresponding to the second and first excited state, respectively, $ A'\,^2\Delta_{5/2}$ and $ A'\,^2\Delta_{3/2}$.
\\
The fluorescence detection of the first excited state, starting from 1000\,nm, is greater than the one shown in Figure \ref{fig:scan2D}, as this measurement is limited by the silicon spectral sensitivity of the spectrometer.
\\
Two additional emission features are detected. The first is excited at approximately 750\,nm, and exhibits a broad fluorescence band centered around 1107\,nm. Its spectral profile is comparable with the one reported in our previous work on neutral barium embedded in a neon matrix, corresponding to an excitation to the 5d6s $^3$D$_1$ state \cite{lippi2025neutralbariumsolidneon}. This confirms the production of multiple species during laser ablation of the BaF$_2$ target. A second feature appears as a broad excitation band extending from approximately 815\,nm to 700\,nm, with an emission maximum centered around 1045\,nm. At present, the origin of this emission remains unclear. Its broad excitation profile and spectral characteristics suggest that it does not originate from isolated BaF molecules, although additional investigations will be required to determine its nature.
\\
Overall, the observed excitation spectrum is consistent with previous studies, indicating that the molecular environment is optically comparable to that obtained in other neon-matrix experiments, despite the lower substrate temperature achieved in the present setup.
\subsection*{Density of molecules}
From the fluorescence signal obtained under resonant excitation of the $X^2\Sigma^+(v=0)$ to $ A^2\Pi_{1/2}(v'=0)$ transition at 859 nm, we estimate the density of BaF molecules in the neon matrix.
\\
As the oscillator strength of the molecules in a matrix remains approximately unchanged, the excitation width varies from a gas phase value $\Delta\nu_{\mathrm{gas}} \simeq 100$~MHz to $\Delta\nu_{\mathrm{mat}} \simeq 2$\,THz in a matrix.
\\
From the calculation shown in Appendix \ref{app}, where the BaF cross section in a neon matrix is estimated, we obtain a density ($n$) times thickness of the crystal ($l$) of BaF molecules during a growth of an hour that is in the order of magnitude of $nl = 5\times 10 ^{13} \; \text{cm}^{-2}$.

\subsection*{Optical bleaching of barium monofluoride in a neon matrix} 
\begin{figure}[]
\centering
\includegraphics[width=1\linewidth]{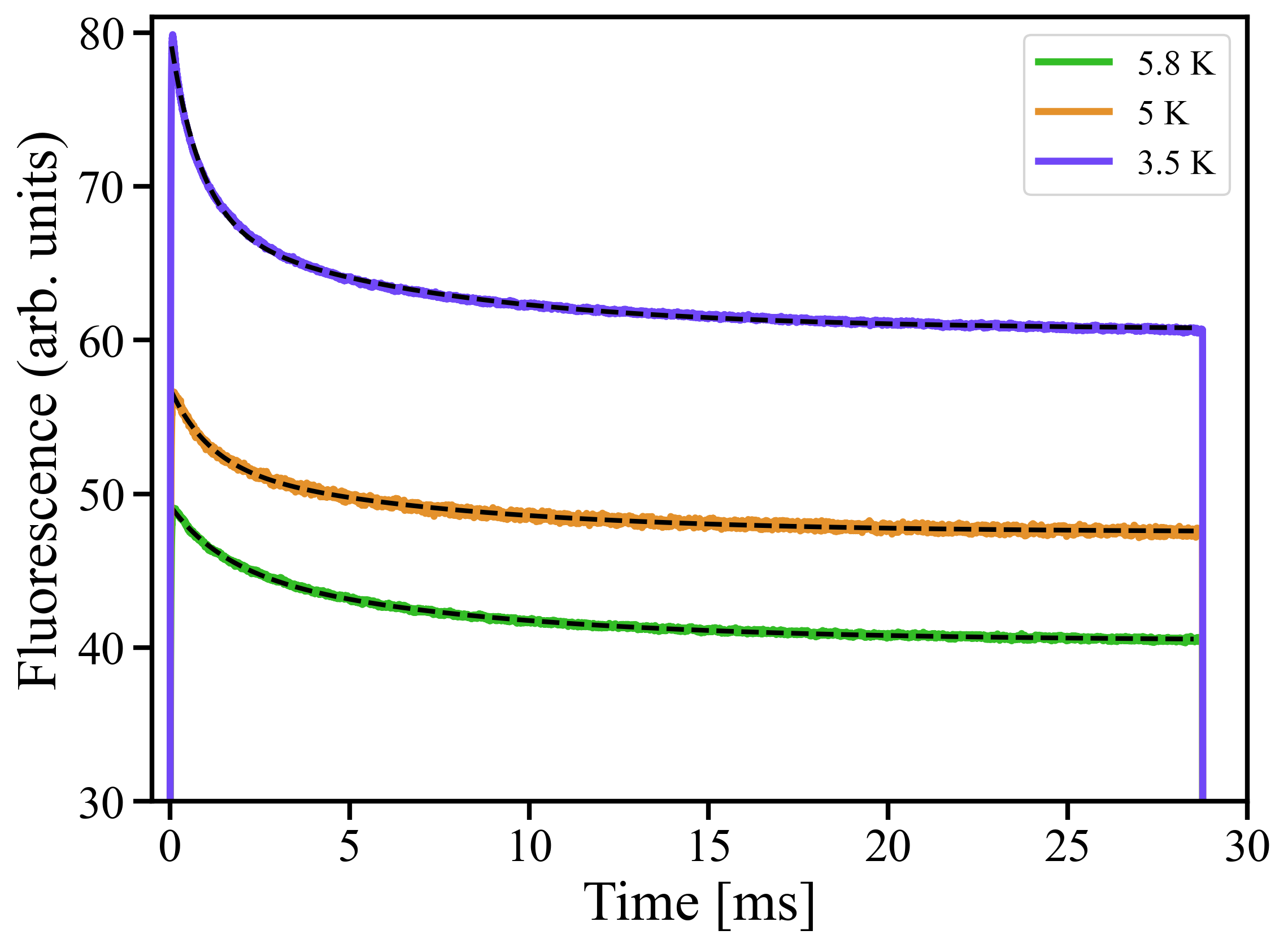}
\caption{Time evolution of the fluorescence signal at different temperatures from the $A'\,^2\Delta_{5/2}$ state of BaF molecules embedded in a neon matrix. Excitation is achieved with 28\,ms long laser pulses at 28\,mW tuned at 859.8\,nm with a period of 1\,s. Lower temperatures yield a higher initial fluorescence intensity but also lead to a more pronounced optical bleaching. The dashed black line indicates the double exponential fit of the decay. Results are shown in Table \ref{tab:temperature_bleaching}. }
\label{fig:bleaching}
\end{figure}
\begin{figure}[]
\centering
\includegraphics[width=1\linewidth]{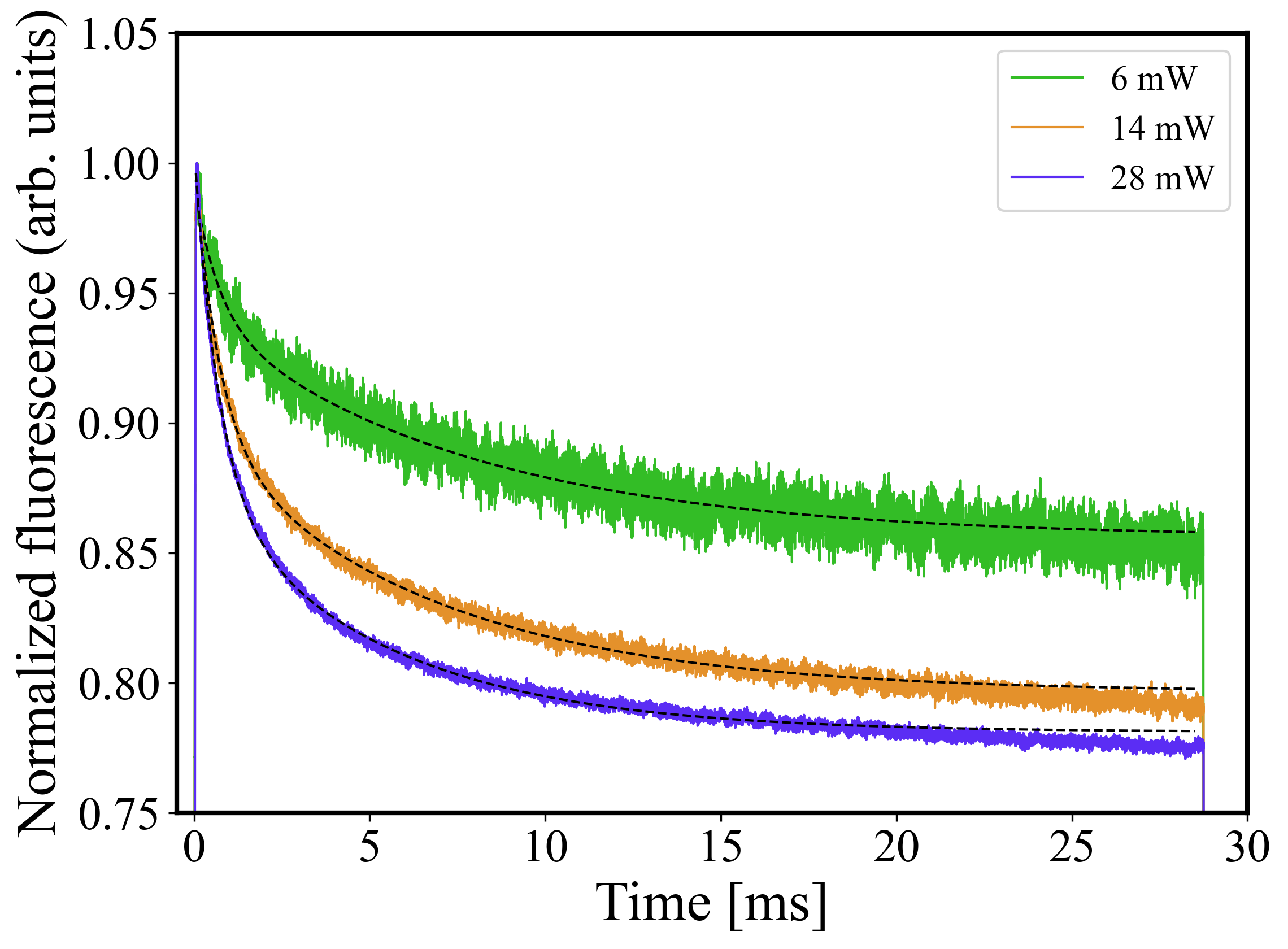}
\caption{Normalized fluorescence fast decay dynamic of the $A'\,^2\Delta_{5/2}$ state of BaF when shined with different laser intensity. The substrate is at 3.5\,K while the excitation is a 28\,ms laser pulse tuned at 859.8\,nm with a period of 1\,s. Increasing the excitation power results in an increased degree of optical bleaching decay. The dashed black line indicates the double exponential fit of the decay. Results are shown in Table \ref{intensity_bleach}.}
\label{fig:bleaching_laser_int}
\end{figure}

\begin{table}[h!]
\centering
\caption{Parameters obtained from the double exponential fit of the fluorescence bleaching decay at different temperatures. The fluorescence signal was fitted with the function $D(t)=a+b\cdot e^{-t/\tau_1}+c\cdot e^{-t/\tau_2}$, where $\tau_1$ and $\tau_2$ describe the slow and fast bleaching components, respectively. The parameter $P$ represents the percentage of the fast component amplitude $c$ relative to the asymptotic level $a$.}
\begin{tabular}{c c c c c c c c}
\hline
T (K) & $a$ (mV) & $b$ (mV) & $\tau_1$ (ms) & $c$ (mV) & $\tau_2$ (ms) & $P$ (\%) & $R^2$ \\
\hline
5.8 & 40.45 & 5.06 & 7.37 & 3.74 & 1.48 & 9.2 & 0.999 \\
5.0 & 47.51 & 4.49 & 7.01 & 4.87 & 1.10 & 10.2 & 0.998\\
3.5 & 60.69 & 6.98 & 6.78 & 12.16 & 0.854 & 20.0 & 0.998 \\
\hline
\end{tabular}
\label{tab:temperature_bleaching}
\end{table}
In an attempt to reproduce the optical pumping scheme proposed by Li et al. \cite{Vutha2023} to populate the dark state $X^2\;\Sigma_{1/2}\;(F = 1,\,m_f = 1)$ using circularly polarized excitation light and a static magnetic field, fast decay of the fluorescence signal has been detected, independent by both polarization and magnetic field, therefore unrelated to the pumping scheme.
\\
This decay occurs on a millisecond timescale from the start of the excitation pulse, as shown in Figure \ref{fig:bleaching}. Although its duration is comparable to the expected time scale of the population pumping scheme, it depends mainly on the substrate temperature and intensity of the laser and recovers after a few seconds in the dark. This behavior could be due to local heating of the sample or other bleaching processes, in agreement with other studies on bleaching effects in noble gas matrices \cite{yvaine2024overcomingBleaching_BaXenon_nEXo, lancaster2021radiative}. 
\\
A series of measurements to investigate the phenomena have been carried out by shining a 28\,ms laser pulse tuned at 859.8\,nm with a period of 1\,s while varying excitation laser intensity or substrate temperature.
\\
A strong influence on bleaching dynamics is observed at lower temperatures, as shown in Figure \ref{fig:bleaching}. This behavior can be consistent with the suppression of thermally activated recovery processes, thus with an increase of the fluorescence light detected. 
\\
 To quantify bleaching dynamics, the decay has been analyzed using a double exponential fit such as $D(t)=a+b\cdot e^{(-t/\tau_1)}+c\cdot e^{(-t/\tau_2)}$, where $a$ indicates the baseline level of the fluorescence signal,  $\tau_1$ and $\tau_2$ describe the slow and fast decay characteristic times, while $b$ and $c$ describe their corresponding amplitudes. The results are shown in Table \ref{tab:temperature_bleaching}. To quantify the bleaching amplitude, we define the parameter $P=100 \times c/a$ as the contribution of the fast decay component relative to the baseline fluorescence level $a$. As the temperature decreases, $P$ increases from 9.2\% at 5.8\,K to 20.0\% at 3.5\,K, indicating that the fast bleaching channel becomes progressively more significant at lower temperatures.
\\
It should be noted that operation at low temperatures is essential for matrix-isolated eEDM experiments, as it reduces the number of phonon interactions, thus increasing the spin coherence time and consequently the measurement sensitivity\,\cite{vutha2018oriented}.
\\
A similar analysis has been conducted related to the intensity of the excitation laser, as shown in Figure \ref{fig:bleaching_laser_int}. The corresponding fit parameters are reported in Table \ref{intensity_bleach}. Here, $P$ increases from 5.8\% to 16.8\% passing from 6\,mW to 28\,mW on the excitation laser, indicating an higher amplitude of the fast bleaching decay channel. Again, high laser power are required to efficiently optical pump the molecular population into the desired hyperfine state for an eEDM measurement, however a local heating of the sample might be happening at higher intensities of laser excitations.
\\
For the 14 and 28 mW data, small deviations from the double exponential fit appear in the last few milliseconds of the pulse, possibly due to a slower bleaching component. The fit uncertainty, evaluated with $\chi^2_{red}$, ranges from 0.3 to 1.4 across the two datasets of power and temperature behavior. This indicates adequate double-exponential description of the decay but also the need for a third exponential for a better fit after 20 ms from the start of the pulse, beyond the time window of interest for this work and is left for future study.
\begin{table}[h!]
\centering
\caption{Parameters obtained from the double exponential fit of the fluorescence bleaching decay at different excitation laser intensities. The fit function $D(t)=a+b\cdot e^{(-t/\tau_1)}+c\cdot e^{(-t/\tau_2)}$ is described in table \ref{tab:temperature_bleaching}. The parameter $P$ represents the percentage of the fast component amplitude $c$ relative to the asymptotic level $a$.}
\begin{tabular}{c c c c c c c c}
\hline
Power (mW) & $a$ (mV) & $b$ (mV) & $\tau_1$ (ms) & $c$ (mV) & $\tau_2$ (ms) & $P$ (\%) & $R^2$ \\
\hline
6  & 14.11 & 1.42 & 7.59 & 0.82 & 0.677 & 5.8 & 0.991 \\
14 & 39.81 & 4.97 & 6.55 & 5.27 & 0.694 & 13.2 & 0.998\\
28 & 62.42 & 7.45 & 5.22 & 10.52 & 0.704 & 16.8 & 0.998 \\
\hline
\end{tabular}
\label{intensity_bleach}
\end{table}

\section{Conclusions}
In this work, we successfully developed a new experimental setup capable of growing cryogenic crystals at temperatures down to 3.2\,K, investigating neon crystals doped with BaF molecules produced by laser ablation in a buffer gas cooling cell since we control the molecular source and the substrate temperature independently. The rotatable substrate, mounted on a movable flange, allows multiple spectroscopic investigations with a single crystal growth.
\\
In the same apparatus, we also demonstrated the production of BaF using p-H$_2$ as a buffer gas, preparing the ground for the growth and spectroscopic investigation of p-H$_2$ crystals doped with BaF in future studies.
\\
Furthermore, BaF fluorescence has been investigated in the spectral region corresponding to the first and second excited electronic states. We identified the previously reported excitation bands \cite{Vutha2023} and extended the excitations to 700 nm. Within this spectral range we also observed fluorescence from atomic barium, which can be distinguished from that of the BaF, demonstrating that laser ablation produces both species. Moreover, we obtained a first estimate of the molecular surface density of BaF in such crystals: $nL = 5 \times 10 ^{13} \; \text{cm}^{-2}$. 
\\
Finally, our study on bleaching dynamics identified temperature and laser intensity as key parameters governing the bleaching phenomena in neon crystals doped with BaF, making bleaching effects an important systematic effect to be studied in future experiments.
\\

\begin{acknowledgments}
We thank E. Berto, F. Calaon, M. Tessaro, M. Rebeschini, and M. Zago for their invaluable technical support.
\\
We express our sincere gratitude to A. C. Vutha and S. Hoekstra for
contributing to discussions that significantly enriched this work.
\end{acknowledgments}

\section*{AUTHOR DECLARATIONS}
\subsection*{Conflict of Interest}
The authors have no conflicts to disclose.

\subsection*{Author contributions}
\textbf{Alessandro Lippi:} Conceptualization (equal); Data curation (lead); Formal analysis (equal); Investigation (equal); Methodology (equal); Software (lead); Visualization (lead); Writing – original draft (lead); Writing – review \& editing (equal).
\\
\textbf{Diego Bisero:} Resources (equal); Validation (equal).\\
\textbf{Caterina Braggio:} Resources (equal); Validation (equal).\\
\textbf{Giovanni Carugno:} Conceptualization (equal); Funding acquisition (lead); Methodology (equal); Project administration (lead); Resources (equal); Supervision (lead); Writing – review \& editing (equal).\\
\textbf{Roberto Calabrese:} Funding acquisition (equal); Resources (equal).\\
\textbf{Federico Chiossi:} Validation (equal); Writing – review \& editing (equal).\\
\textbf{Marco Guarise:} Formal analysis (equal); Funding acquisition (equal); Resources (equal); Supervision (equal); Validation (equal); Writing – review \& editing (equal).\\
\textbf{Madiha M. Makhdoom:} Investigation (equal); Writing – review \& editing (equal).\\
\textbf{Giuseppe Messineo:} Conceptualization (equal); Investigation (equal); Methodology (equal); Validation (equal); Writing – review \& editing (equal).\\
\textbf{Jacopo Pazzini:} Funding acquisition (equal); Resources (equal); Writing – review \& editing (equal).\\

\subsection*{Funding}
This research was funded by the PNRR MUR project PE0000023 – NQSTI, financed by the European Union—Next Generation EU. This work was also supported in part by the Italian MUR “Departments of Excellence grant 2023–2027—Quantum Frontiers”, by the “Progetti di Ricerca di Rilevante Interesse Nazionale” (PRIN grant 20227F5W4N, PRIN 2022), and by INFN CSNV within the framework of DOCET experiment.

\section*{Data availability}
The data supporting the findings of this study are available
from the corresponding author upon reasonable request.


\clearpage

\appendix

\section{Estimate of the BaF density from fluorescence measurements}
\label{app}

An estimate of the absorption cross sections ($\sigma_{\text{abs}}$) at 859 nm and 815 nm is here proposed.
\\
The cross section is the ratio between the transition rate $W_{eg}$ of the interested states X$^2$$\Sigma^+$ $v=0$ and the excited state, either A $^2\Pi_{1/2}$ $v'=0$ (859 nm) or A $^2\Pi_{3/2}$ $v'=0$  (815 nm), and the photon flux $\phi$ as:
\begin{equation}
    \sigma_{abs} = W_{eg}/\phi \,.
\end{equation}
Here:
\begin{equation}
    W_{eg} = \frac{\epsilon_0^2 \gamma_p c^3 (2J' +1) }{4 \hbar \gamma_{tot} \omega_0^3 (2J+1)} 
\end{equation}
where $ \gamma_{tot}$ and $\gamma_{p}$ are respectively the total and partial linewidth of the investigated transition, defined as:
\begin{align}
    &\gamma_{tot} = \frac{2 \pi}{\lambda} \sqrt{\frac{8 k_B T log 2}{m}}
    \\   
    &\gamma_{p, eg} = \epsilon_{eg} \gamma 
\end{align}
where $\gamma$ is the natural linewidth and $\epsilon_{eg}$ is the branching ratio:
\begin{equation}
    \epsilon_{eg} = F_{v(0,0)} \frac{H_{eg}}{\sum_i H_{ei}}
\end{equation}
where $F_{v(0,0)}$ is the Franck-Condon factor for transition from $v=0$ to $v' =0$ and $H_{ei}$ is the Honl-London factor for the $e \xrightarrow{}g$.
\\
Given a transition from a $\Sigma$ ground state to a $\Pi$ excited state \cite{herzberg1950spectra} and an absorption ruled by Doppler broadening \cite{Budker2008}, the absorption cross section for a transition $J' = J+1$ can be computed as:
\begin{equation}
    \sigma_{abs} = \frac{\lambda^3}{2(2\pi)^2}\frac{(2J+3)(J+2)}{(2J+1)^ 2 } \sqrt{\frac{m}{8k_BTlog2 }} \frac{F_{v(0,0)}}{\tau}
\end{equation}
where $\tau$ represents the lifetime of the investigated state, which is $\tau = 56.0$ ns \cite{Berg1998} for the $A\,^2\Pi_{1/2}$ state  (or 57.1 ns \cite{Aggarwal2019}) and 47.9 ns for the $A\,^2\Pi_{3/2}$ state \cite{Aggarwal2019}.
\\
Considering molecules cooled to a buffer gas temperature $T_{neon} = 20$ K, one gets:
\begin{align}
    \sigma_{abs}(859) = 1.40 \times10^{-11} \; \text{cm}^2
    \\
    \sigma_{abs}(815) = 1.32 \times10^{-11} \; \text{cm}^2 \,.
\end{align}

\subsection*{Absorption cross section in the matrix}
To compute the cross section in a cryogenic matrix environment, we start from the oscillator strength (interaction strength) $F$ of the optical transition. When moving from a gas phase to a neon matrix, $F$ is expected to remain approximately unchanged, as the local field correction is small and given by:
\begin{equation}
F_{\text{mat}} \approx F_{\text{gas}} \left( \frac{\eta^2 + 2}{3} \right)^2
\end{equation}
For a refractive index $\eta \sim 1.1$, this corresponds to a correction below $\sim 20\%$. Additionally, crystal field mixing in the neon matrix is expected to be weak. This is consistent with the fact that radiative lifetimes in gas and matrix differ only slightly.
\\
The relation between absorption cross section and oscillator strength is:
\begin{equation}
\int \sigma_{\text{gas}}(\nu)\, d\nu = \frac{\pi e^2}{m_e c} F_{\text{gas}} \approx \frac{\pi e^2}{m_e c} F_{\text{mat}} = \int \sigma_{\text{mat}}(\nu)\, d\nu
\end{equation}

Defining a characteristic linewidth $\Delta \nu = \int d\nu$, we obtain peak cross section scaling as:
\begin{equation}
    \sigma_{\mathrm{peak,Ne:BaF}} \approx \sigma_{\mathrm{peak,gas}}\,\frac{\Delta\nu_{\mathrm{gas}}}{\Delta\nu_{\mathrm{mat}}}.
\end{equation}
Using the previously obtained values: 
$ \sigma_{\mathrm{peak,gas}} \simeq 1.4\times10^{-11}\;\mathrm{cm^2}$,
$\Delta\nu_{\mathrm{gas}} \simeq 100$ MHz, and $\Delta\nu_{\mathrm{mat}} \simeq 2$ THz, one gets a cross section in a neon matrix environment of BaF molecules of:
\begin{equation}
    \sigma_{\mathrm{peak,Ne:BaF}} \simeq 7\times10^{-16}\;\mathrm{cm^2}.
\end{equation}

\subsection*{Number of emitted photons} 

The number of absorbed photons can be estimated using the Beer–Lambert law in the weak absorption limit ($n \sigma_{\text{peak, Ne:BaF}} L \ll 1$), where :
\begin{align}
N_{\text{abs}} = N_{\text{in}} \left(1 - e^{-n \sigma_{\text{peak, Ne:BaF}}(\nu_0) L} \right)
\\
\approx N_{\text{in}} \cdot n \cdot \sigma_{\text{peak, Ne:baF}}(\nu_0) \cdot L
\label{18}
\end{align}
where $n$ is the density of BaF molecules, $N_{\text{in}}$ is the number of incident photons and $L$ is the matrix thickness.
\\
If the excited molecules decay radiatively with quantum yield (branching ratio) $\Phi$, the number of detected photons is
\begin{equation}
    N_{\mathrm{det}} = N_{\mathrm{abs}}\,\Phi\,\frac{d\Omega}{\Omega},
    \label{19}
\end{equation}
where $d\Omega/\Omega$ is the collection efficiency of the detector.

\subsection*{Collected photons}

With a silicon detector with responsivity $R = 0.6$ A/W at 950-970 and a transimpedance gain $G = 5\times10^4$ V/A, a measured signal of $V = 14$ mV in the spectral region of the $ A'\,^2\Delta_{5/2}$ state (940 - 970 nm), over a time window $\Delta T = 1$ ms, corresponds to a number of detected photons of:
\begin{equation}
    N_{\mathrm{det}} = \frac{V}{G}\,\frac{1}{R}\,\frac{1}{E_{\mathrm{ph}}}\,\Delta T \simeq 2.7\times10^{9}.
\end{equation}
At $\lambda = 860$ nm and $P = 40$ mW, the number of incident photons in $1$ ms is
\begin{equation}
    N_{\mathrm{in}} \simeq 1.7\times10^{14}.
\end{equation}
Hence,
\begin{equation}
    \frac{N_{\mathrm{det}}}{N_{\mathrm{in}}} \simeq 1.6\times10^{-5}.
\end{equation}

\subsection*{Final estimate for the matrix density}

Using the matrix-broadened cross section derived from the gas-phase value $ \sigma_{\mathrm{peak,Ne:BaF}} \simeq 7\times10^{-16}\;\mathrm{cm^2}$ we obtain using eq. \ref{18} and \ref{19}:
\begin{align}
    nL =& \frac{N_{\mathrm{det}}}{N_{\mathrm{in}}}\,\frac{1}{\sigma_{\mathrm{mat}}}\,\frac{1}{\Phi\,d\Omega/\Omega} 
    \\
    =& \frac{1.6 \times10^{-5}}{7 \times10^{-16}}\frac{1}{\Phi\,d\Omega/\Omega}
\end{align}
Assuming $\frac{d\Omega}{\Omega} \sim 0.05$ and a branching ratio of $\Phi = 10^{-2}$ (it is $3 \times 10^{-4}$ in the gas phase \cite{BranchingratioBaFHao_2019}, however, non radiative pathways can  lead to a higher BR), we find:
\begin{equation}
    nL \sim 5 \times10^{13}\;\mathrm{cm^{-2}}.
\end{equation}

\newpage

\bibliography{bibliography}

\end{document}